# Spin-lattice coupling mediated giant magnetodielectricity across the spin reorientation in Ca$_2$FeCoO$_5$

Gaurav Sharma[1], Shekhar Tyagi[1], V. R. Reddy[1], A.M. Awasthi[1], R.J. Choudhary[1], A.K. Sinha[2] and Vasant Sathe[1*]

[1] UGC-DAE Consortium for Scientific Research, D. A. University Campus, Khandwa Road, Indore-452001, India
[2] Indus Synchrotrons Utilization Division, Raja Ramanna Centre for Advanced Technology, Indore-452013, India

*Corresponding Author: vasant@csr.res.in

**Abstract:** The structural, phonon, magnetic, dielectric, and magneto dielectric responses of the pure bulk Brownmillerite compound Ca$_2$FeCoO$_5$ are reported. This compound showed giant magneto dielectric response (10%–24%) induced by strong spin-lattice coupling across its spin reorientation transition (150–250 K). The role of two Debye temperatures pertaining to differently coordinated sites in the dielectric relaxations is established. The positive giant magneto-dielectricity is shown to be a direct consequence of the modulations in the lattice degrees of freedom through applied external field across the spin reorientation transition. Our study illustrates novel control of magneto-dielectricity by tuning the spin reorientation transition in a material that possess strong spin lattice coupling.

**Introduction**

Magneto-dielectric compounds hold great promise due to potential applications in futuristic devices [1,2]. In such compounds, the mutual effective control of electrical and magnetic properties holds the key for promising applications. Number of materials showing such effects are scarce due to the mutual exclusion of spontaneous electrical dipolar order and spin order for electronic reasons; the essentiality for magnetic ordering is partially filled d-bands which hinders the dipolar ordering [3,4]. In order to circumvent this condition, compounds showing spiral spin ordering are thought to be promising candidate as the spiral spin order destroys locally the centro-symmetry of the ions enabling polarization. However, the magneto dielectric or magneto-electric effect in such compounds is found to be very weak, barring few compounds [5,6]. The incommensurate spiral spin arrangement is suggested to be the root cause for the induced polarization in CuO [7], while, the charge ordering of Fe$^{2+}$ and Fe$^{3+}$ is suggested to be responsible for induced ferroelectricity in LaFe$_2$O$_4$ [8]. On the other hand geometric frustration is attributed to the improper ferroelectricity observed in Yttrium manganites [9]. In partially ordered double perovskite La$_2$NiMnO$_6$, it is shown that the anti-site disorder of the cations generates significant asymmetric hopping under magnetic field resulting in giant magneto dielectric effect at room temperature [10]. In most of these compounds, negative giant magneto dielectric effects are shown. A positive giant magneto dielectric effect was reported in TbMnO$_3$ single crystal at low temperatures [11]. Here, the frustrated sinusoidal antiferromagnetic order induced magneto-elastic behaviour was attributed for the induced polarization. In this compound, it was suggested that the spin reorientation of Mn$^{3+}$ caused by the plausible magnetic field induced Tb$^{3+}$ moment reversal changes the exchange interaction energy and then brings about the lattice modulation owing to a finite spontaneous polarization. However, no direct evidence of spin reorientation of Mn$^{3+}$ was provided. In the spin reorientation transition (SRT) region, the applied magnetic field is expected to induce frustration leading to the induced polarization.

Ca$_2$FeCoO$_5$ is a Brownmillerite [12,13] type compound with orthorhombic crystal structure in *Pbcm* space group with unit cell parameters, $a$=5.3626(6), $b$=11.0943(4) and $c$=14.8109(6) [14]. The fact that one of the short lattice parameters is doubled makes this compound rare among the Brownmillerite compounds with a supercell twice the size of a regular Brownmillerite unit cell. The formation of super structure causes the formation of two sets of octahedral and tetrahedral sites. This compound also exhibits intra-layer cation ordering which is rare, even among Brownmillerite compounds. The tetrahedral sites exhibit complete Fe/Co ordering while the octahedral sites have certain degree of randomness [14,15]. It must be noted that the compound exhibits an overall G-type anti-ferromagnetic order with tetrahedral and octahedral sites exhibiting anti-ferromagnetic order individually with different ordering temperatures. Neutron diffraction studies on this compound reveals that the spin easy axis in this compound changes from along the *b* axis below 100 k to along the shortest axis above 200 K through a broad spin reorientation transition [14].

Here, we report the direct evidence of strong spin-lattice coupling across the spin reorientation transition and huge magneto-dielectric coupling in Ca$_2$FeCoO$_5$ compound. The compound is probed in detail by temperature dependent magnetization, synchrotron x-ray diffraction (SXPD), Raman spectroscopy, Mossbauer Spectroscopy, and complex dielectric measurements. Most importantly, the maximum value of magneto dielectricity obtained was ~24% at the temperature value ~220K for the frequency of ~5 kHz, making it worthy of industrial applications.

**Experimental Details**

The sample was synthesized by solid state reaction method, using high purity CaCO$_3$ (99.99%), Fe$_2$O$_3$ (99.99%), Co$_3$O$_4$ (99.99%) as precursors. Pellets were made after number of intermediate heating to the powder and were sintered at 1250°C for 33 hours. The synchrotron x-ray diffraction (SXRD) data was collected using a He CCR installed at BL12 Indus-II synchrotron source, RRCAT, India. The wavelength used was 0.782566 Å. The data was fitted by Rietveld refinement method [16] using FullProf software [17], the fitted pattern is shown in **Figure 1** and the resulting lattice parameters are, $a$=5.3626(6), $b$=11.0940(9), $c$=14.8109(4). The refinement results are in accordance with the previous report [14].

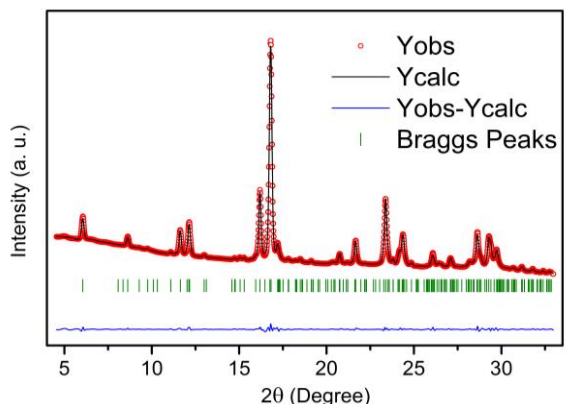

**Figure 1**: Room temperature synchrotron x-ray diffraction pattern of $Ca_2FeCoO_5$ with the Reitveld refinement profile along with the difference pattern and Braggs peaks. The wavelength used was 0.782566 Å, the space group used during refinement was Pbcm,

The Raman spectroscopy studies were carried out using LabRam HR800 System, equipped with a 473 nm excitation source, an 1800g/mm grating and THMS 600 temperature variation stage from Linkam, U. K. The magnetization measurements were carried out in zero field cooled (ZFC), field cooled cooling (FCC) and Field cooled warming (FCW) protocols in 20 Oe applied field using superconducting quantum interface device vibrating sample magnetometer (SQUID-VSM) from Quantum Design Inc., USA.

The $^{57}$Fe Mossbauer measurements as a function of temperature were done in transmission mode with $^{57}$Co/Rh radioactive source in constant acceleration mode with Wissel velocity drive. Velocity calibration was done using natural Fe. Low temperature and high magnetic field Mossbauer study was carried out using a Janis make superconducting magnet with 5 Tesla magnetic field. The dielectric measurements with parallel plate capacitor arrangement over 1kHz to 100 kHz were performed using Alpha-A broadband impedance analyser from Novo Control. Magnetic field and temperature dependent complex dielectric measurements from 6K to 300 K were performed using Oxford Nanosystems Integra 9T magnet-cryostat.

**Results and discussions**

The crystal structure of this compound is unique even in Brownmillerites. The $b$-axis is doubled due to intra layer cationic ordering. Thus, this compound supports both intra- and interlayer cation site ordering. On top of it, Brownmillerites are known to show ordering of oxygen vacancies. Thus, determining the valance state of the two cations and site order is important. In order to verify the oxidation states of the two cations i.e. Fe and Co, temperature dependent x-ray absorption near edge spectroscopy (XANES) was carried out at the Fe and Co K-edge that is shown in the **Figure 2**. The overall spectra and edge position of the two cations matched with the previous report [18] confirming 3+ valence state of the two cations. For comparison, the XANES spectra of $Fe_2O_3$ standard sample is also plotted in the inset of the figure along with the zoomed view across the Fe K-edge. The edge position matches exactly in the two samples confirming the 3+ valance state of the Fe cations. It is also observed that the edge position is invariant with temperature.

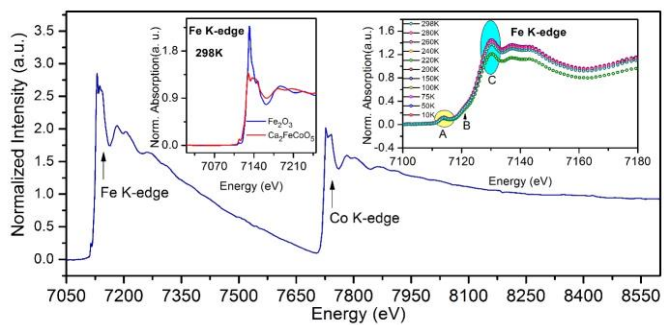

**Figure 2**: XAS data of $Ca_2FeCoO_5$ at Fe and Co K-edge. The first inset shows the Fe k-edge XANES data of $Ca_2FeCoO_5$ along with the reference $Fe_2O_3$ collected at room temperature showing that the oxidation state of Fe in the sample is +3. The second inset shows the temperatures variation of the Fe K-edge XANES indicating that the valence of the cation remains unchanged as the temperature is varied.

The magnetic structure of this compound was reported to be G-type anti-ferromagnetic with Neel temperature above 450 K. Most importantly, neutron diffraction and magnetization studies showed the presence of a spin reorientation transition around 200 K. In order to verify these findings, the magnetization studies as a function of temperature were carried out and are presented in **Figure 3**.

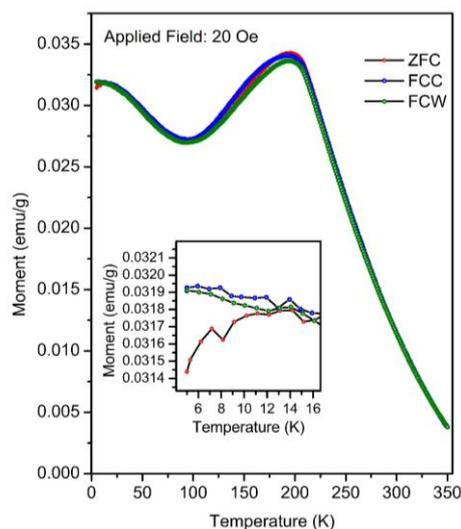

**Figure 3**: Temperature dependent magnetization (M-T) of $Ca_2FeCoO_5$ from 5K to 350K measured in Zero-Field Cooled (ZFC), Field Cooled Cooling (FCC) and Field Cooled Warming (FCW) protocols. The inset shows the zoomed view of low temperature region. The increasing moment below 100K in all the curves also point towards the presence of some degree of spin canting in this system.

The slope changes at ~200K and ~100K are consistent with reported spin re-orientation transition (SRT) [14]. In order to investigate the magnetic structure in detail, $^{57}$Fe Mossbauer spectroscopy as a function of temperature was carried out from 5 K to 300 K. The Mossbauer spectrum collected at different temperatures are shown in **Figure 4** (a) to (h). As mentioned before, the $Fe^{3+}$ ion occupies the octahedral as well as tetrahedral sites in this compounds that independently exhibit anti-ferromagnetic order. Therefore, the spectrum were fitted with two broad sextets representing octahedral and tetrahedral Fe ions. The obtained isomer shift values confirm the presence of octahedrally and tatrahedrally coordinated $Fe^{3+}$. The presence of sextet at 300K is consistent with the antiferromagnetic order at room temperature and the overall occupation of Fe at octahedral and tetrahedral sites were found to be 50.9% and 49.1%, respectively. The observation of two broad sextets with an effective field equal to about $(H_{int} \pm H_{ext})^{1/2}$ and the presence of intense $\Delta m=0$ lines in the in-field Mossbauer data (5K|5Tesla) {**Figure 5**} confirms the anti-ferromagnetic order in the sample.

from 300 to 5 K for these two magnetic components at the SRT {**Figure 4** (i)}, which is a consequence of the change in the direction of component of electric field gradient parallel to the internal magnetic field direction with the spin reorientation transition [19,20, 21].

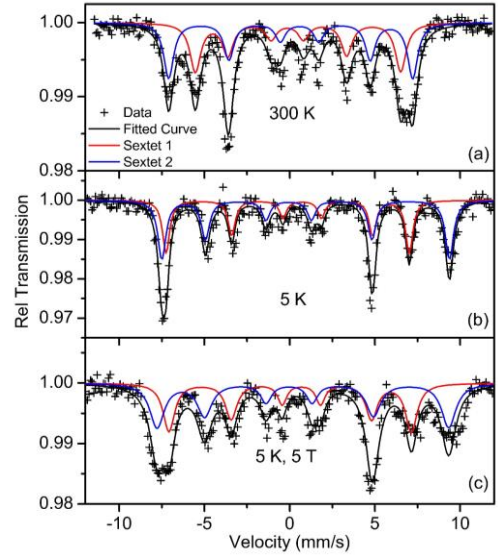

**Figure 5**: Mossbauer spectroscopy data collected at 300 K (a), collected at 5 K (b) and at 5 K under the application of 5 Tesla magnetic field (c) along with the fitting considering the spectrum to be a convolution of two broad sextets representing differently coordinated sub-lattices, namely octahedral and tetrahedral.

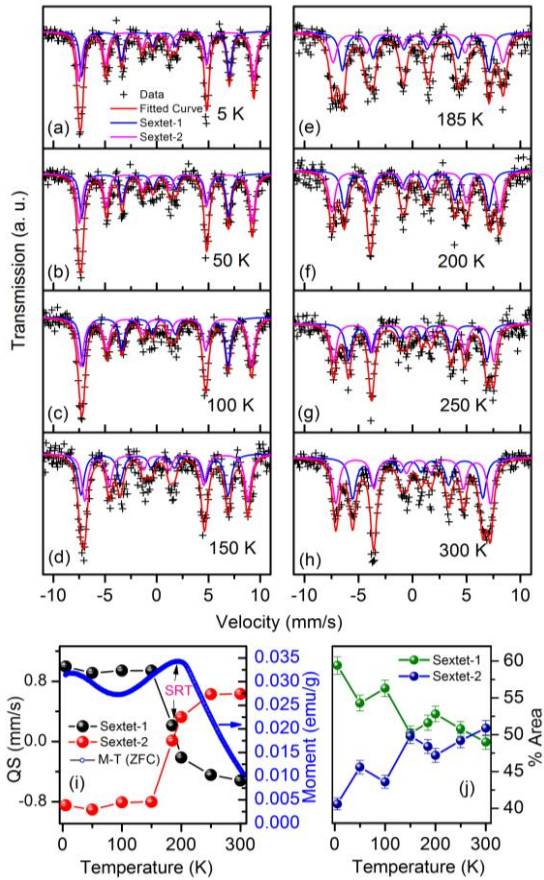

**Figure 4**: Temperature dependent Mossbauer measurements carried out from 300K to 5K (a-h). Panel (i) shows the quadrupole splitting for $Fe_T$ and $Fe_O$ sites as a function of temperature. ZFC magnetization is also shown to mark the SRT. Panel (j) shown the % area of the overall pattern covered by different sextets representing the Fe occupation at $Fe_T$ and $Fe_O$ sites.

A detailed analysis of the Mossbauer spectra resulted in various values of FWHM and hyperfine field etc. that are presented in **Figure 6**. The temperature evolution of quadrupole splitting and the area under the sextets representing the Fe ions at octahedral ($Fe_O$) and tetrahedral ($Fe_T$) sites showed some very interesting behaviour, the same is presented in **Figure 4**. It must be noted that the quadrupole splitting values changed sign on lowering the temperature

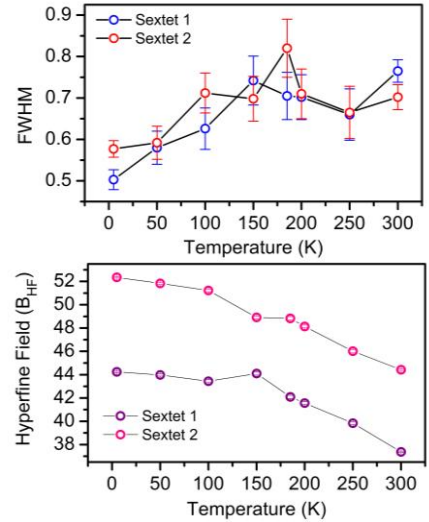

**Figure 6**: The FWHM (a), Isomer Shift (b) and Hyperfine field (c) as a function of temperature. These parameters are obtained by de-convoluting the $^{57}Fe$ Mossbauer spectra of $Ca_2FeCoO_5$ collected at different temperatures. The spectra were fitted using two broad sextets representing octahedral and tetrahedral coordination of Fe in the lattice.

In $Ca_2Fe_2O_5$ the ratio of $Fe^{3+}$ distribution in tetrahedral to octahedral sites ($Fe_T/Fe_O$) is reported to be close to unity, determined from the relative areas of corresponding sextets considering $f_O/f_T$ at 5K as 0.96±0.02 (f- is the recoil free fraction), however, with the doping of Ga, Sc etc., at the Fe site the ($Fe_T/Fe_O$) ratio is found to deviate from unity [22]. In the present work, with Co doping at Fe site, the relative area ratio of $Fe_T/Fe_O$ is found to be about 0.7 at 5 K. Interestingly, the relative area of the two sextets also showed a noticeable change with temperature {**Figure 4** (j)}.

Such, a change is striking and can be explained by considering two possibilities; (1) the relative population of Fe cations at the octahedral and tetrahedral site is changing with temperature or (2) the recoil free fraction of the Fe cations at the two sites is changing with change in temperature around SRT. The first possibility is energetically very costly and hence discarded. The area of each sextuplet is directly related to the number of atoms in a particular coordination and the recoil free fraction for the corresponding Fe nucleus.

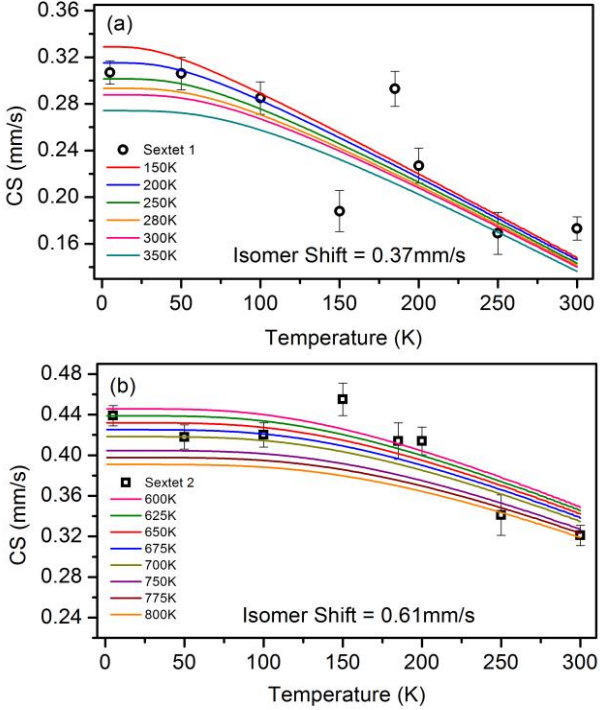

**Figure 7**: Theoretically simulated lines of CSs for given Θ (solid lines) and experimental data (points) for $Fe_T$ (a) and $Fe_O$ (b).

The recoil free fraction $f$ is a function of the mean square displacement of each atom [19], $f = exp\{(-E_R)^2 \langle x^2 \rangle / \hbar c^2\}$. The most typical way to model the $\langle x^2 \rangle$ dependence is through the Debye model, which when applied to the recoil free fraction yields:

$$f = exp\left[-\frac{3E_\gamma^2}{4Mc^2 k_B \Theta}\left(1 + \frac{4T^2}{\Theta^2}\int_0^{\frac{\Theta}{T}} \frac{x\,dx}{e^x - 1}\right)\right] \quad (1)$$

where $E_\gamma$ is the energy of the $^{57}Fe$ gamma ray (14.4 keV), $k_B$ is the Boltzman's constant, M is the mass of $^{57}Fe$ nucleus, c is the speed of light and Θ is the Debye temperature for a particular nucleus. The present results can be explained only by considering drastically different Debye temperatures for the two sites. Such a variation in Θ for two sub-lattices is reported for spinel ferrites, garnets etc [23,24,25,26]. The Debye temperature is directly related with the second order Doppler shift (SODS) [27,28,29]. The SODS is related with the center shift (CS) by the following relation:

$$CS(\Theta,T) = IS + SODS(\Theta,T) \quad (2)$$

here the IS is the isomer shift which is nearly temperature independent while the SODS is defined as:

$$SODS = -\frac{3k_B\Theta}{2Mc}\left[\frac{3}{8} + \frac{3T^4}{\Theta^4}\int_0^{\frac{\Theta}{T}} \frac{x^3 dx}{e^x - 1}\right] \quad (3)$$

We have deduce the experimental values of the CS from our Mossbauer data which is presented in **Table 1**. Using these values and equation 2 and 3, we attempted to calculate the Debye temperatures for the $Fe_O$ and $Fe_T$ sites. A program was constructed, which Debye temperature and an IS would be simulated, resulting in theoretical SODS and CS. As given in reference [20,29], the IS values for the $Fe_O$ was taken as 0.61 mm/s and the variation of CS values for different Debye temperatures were simulated. Similarly, attempts were made to estimate Debye temperature for $Fe_T$ using various values of IS and Θ. The results of this simulation is presented in **Figure 7**. The Debye temperature thus estimated is found to be nearly 700±50 K for the $Fe_O$ while 225±50 K for the $Fe_T$. These values are comparable to the values reported by Kim et al in $CoFe_2O_4$ [24]. The experimental points followed the theoretical trend except in the spin reorientation transition region where it showed dramatic variation in the CS values. This variation is directly related with the variation in SODS which is a signature of lattice dynamical instability in the SRT region. Further, temperature dependent Raman spectroscopy and SXRD studies ascertain the presence of strong spin-lattice coupling across the SRT. The evolution of Raman spectra as a function of temperature is shown in **Figure 8**.

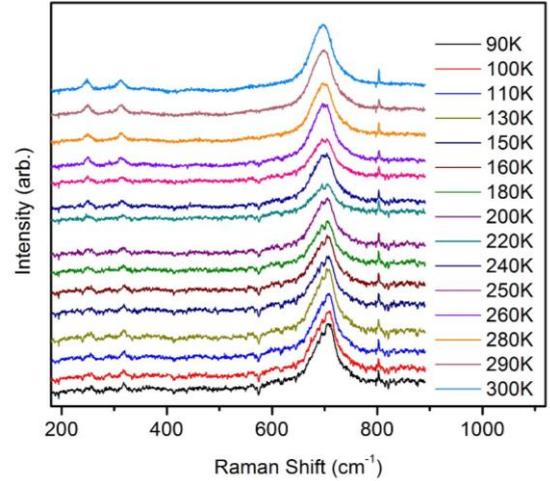

**Figure 8**: Temperature dependence of the Raman Spectrum of $Ca_2FeCoO_5$.

The major mode observed at ~700 cm$^{-1}$ was deconvoluted using two Lorentz functions. The Raman shifts and FWHM thus resulted are plotted as a function of temperature and presented in **Figure 9** (a-b). The Raman shift and FWHM showed anomalous behaviour around the temperature window concurrent with the SRT. The lattice parameters deduced from SXRD is presented in **Figure 9** (c), (d), (e) and (f). The SXRD measurements were carried out in both heating and cooling cycle that showed a strong hysteresis. Such a hysteretic behaviour was also observed in the magnetization measurements (**Figure 3**) confirming metastable behaviour of magnetization and lattice across the SRT.

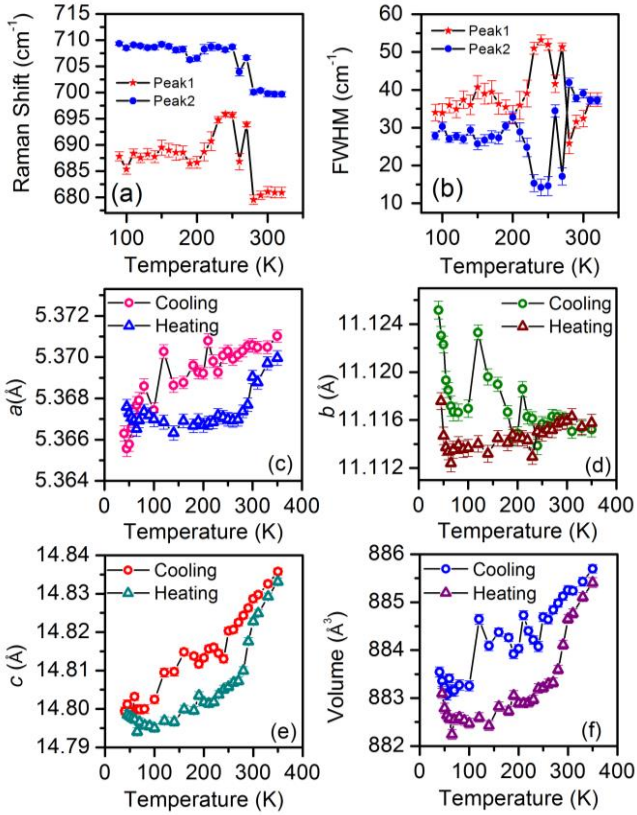

**Figure 9**: The Raman Shift (a) and FWHM (b) of the most prominent modes as a function of temperature and the lattice parameters (c), (d), (e) as a function of temperature obtained from the refinement of SXPD data of $Co_2FeCoO_5$ collected in cooling and heating cycle. The unit cell volume is shown in panel (f).

The dielectric measurements were carried out from 300 K down to 6 K which showed strong frequency dependent dielectric relaxation, however, the dielectric constant is rather small ($\varepsilon'<50$), slowly varying and nearly frequency independent below 100 K (**Figure 10**). The temperature dependent dielectric constant showed two broad humps which are frequency dependent (dispersive). The corresponding loss (tanδ) also showed signatures of strong dielectric relaxations (**Figure 11**). Generally, Maxwell-Wagner processes are considered for explaining extrinsic effects in dielectric relaxations which arises from the grain boundaries. We attempted to measure magneto resistance of the sample in order to detect the extrinsic contribution to the dielectric relaxations, however, the resistance was found extremely high below 250 K and was beyond our measurable limit (>MΩ). Thus, the sample is highly insulating and hence, the contribution of Maxwell-Wagner type relaxation, if at all present, is considered to be negligible. This is corroborated by the low (<100) intrinsic values of the dielectric constant and low (<1) loss tangent at low frequencies, up to 150K. Therefore, the relaxations observed at two distinct temperatures are considered as intrinsic effect of the sample. Mossbauer spectroscopy showed that the Debye temperature corresponding to Fe ions at octahedral and tetrahedral sites are drastically different. Therefore, it is likely that the dielectric relaxations arising due to the two sites are apart in temperature. The two humps in the dielectric measurements are thus attributed to the two Debye temperatures. **Figure 12** (a) shows the typical dielectric constant ($\varepsilon'$) measured at ~23.7 kHz in the SRT region under 0T and 6T applied magnetic field. The insets show derivative of the $\varepsilon'$ collected under 0 T and 6 T, which shows anomaly in the SRT region.

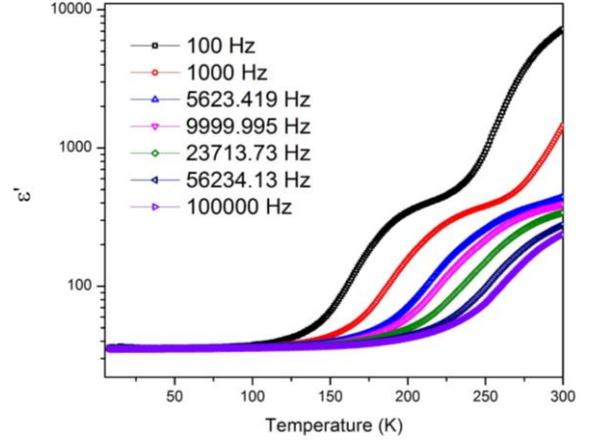

**Figure 10**: The dielectric constant ($\varepsilon'$) as a function of temperature for different frequencies showing the presence of frequency dependent relaxations.

The $\varepsilon'$ measured under 0 and 6T applied magnetic field at 10 kHz and 100 kHz frequencies are shown in **Figure 12** (b) and (d). It clearly shows an enhancement in dielectric constant under field, over the temperature window across the SRT. The corresponding loss tangents are presented in **Figure 12** (c) and (e). It also shows changes in the SRT region. This enhancement is observed at almost all the frequencies, indicating the intrinsic nature of magneto-dielectric effect.

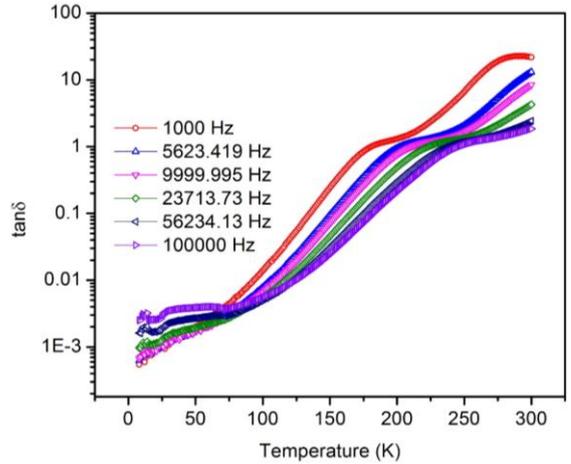

**Figure 11**: The loss tangent (tanδ) as a function of temperature for different frequencies. The presence of two relaxations is very clear from the figure. The relaxation peaks also show a frequency dependence.

The percentage magneto dielectricity (MD) is calculated using formula: MD(%)=[{$\varepsilon'$(6T) −$\varepsilon'$(0T)}/$\varepsilon'$(0T)]×100 and presented in **Figure 4** (f), whereas, the percentage magneto loss (ML) is calculated using formula: ML(%)=[{tanδ(6T)−tanδ(0T)}/tanδ(0T)]×100 for different frequencies. It is observed that the MD is unusually large in magnitude and shows a maximum value of 24% at ~5 kHz which tends to decrease with increasing frequencies, however, the MD is still >10% at 100 kHz. The peak in MD shows a shift towards higher temperature with increasing frequency, reflecting the dispersion observed in $\varepsilon'$.

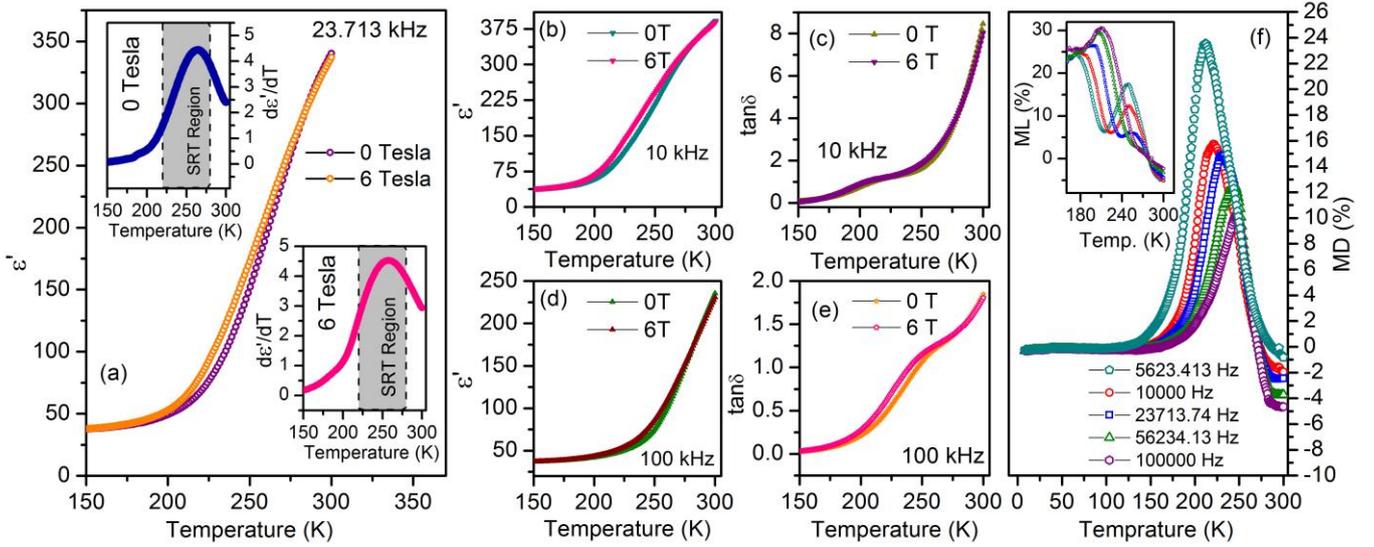

**Figure 12**: The dielectric constant (ε′) vs temperature at 23.713 kHz measured under 0T and 6T applied field, panel (a). The upper (lower) inset shows the derivative of ε′ as a function of temperature collected in 0T (6 T) showing a peak in the SRT region. Panel (b & d) show the dielectric constant (ε′) vs Temperature at typical frequencies of 10 kHz and 100 kHz measured under 0T and 6T, whereas, panel (c & e) shows the corresponding loss tangent as a function of temperature. Panel (f) shows MD (%) calculated from the ε′ vs temperature data under 0 T and 6 T applied field collected at different frequencies, showing a maximum value of 24% for 5623.413 Hz, whereas the inset shows the corresponding ML (%).

**Table 1:** The hyperfine parameters obtained after fitting the Mossbauer spectrum collected at different temperatures where 5 K data was collected with application of 5 Tesla magnetic field as well. The data is fitted with two broad sextets representing octahedral and tetrahedral sites.

| Temperature | FWHM (mm/s) | IS (mm/s) | QS (mm/s) | BHF (Tesla) | Area |
|---|---|---|---|---|---|
| 5 K | 0.50 ± 0.03 | 0.31 ± 0.01 | -0.85 ± 0.02 | 44.3 ± 0.1 | 40.6 |
|  | 0.57 ± 0.02 | 0.44 ± 0.01 | 0.99 ± 0.04 | 52.3 ± 0.1 | 59.4 |
| 5 K, 5T | 0.64 ± 0.03 | 0.36 ± 0.01 | -0.66 ± 0.03 | 44.2 ± 0.2 | 41.6 |
|  | 0.78 ± 0.04 | 0.37 ± 0.01 | 0.81 ± 0.03 | 53.1 ± 0.2 | 58.3 |
| 50 K | 0.58 ± 0.04 | 0.306±0.014 | -0.907±0.03 | 43.98 ± 0.1 | 45.6 |
|  | 0.592 ± 0.04 | 0.418±0.0119 | 0.9098±0.024 | 51.833±0.085 | 54.3 |
| 100 K | 0.626 ± 0.05 | 0.285±0.014 | -0.81±0.033 | 43.44 ± 0.13 | 43.6 |
|  | 0.712±0.048 | 0.42 ± 0.012 | 0.943 ± 0.028 | 51.23 ± 0.119 | 56.3 |
| 150 K | 0.742±0.059 | 0.188±0.0176 | 0.803±0.038 | 44.097±0.13 | 49.8 |
|  | 0.698±0.054 | 0.455 ± 0.016 | 0.943 ± 0.03 | 48.91 ± 0.12 | 50.2 |
| 185 K | 0.705±0.057 | 0.293 ± 0.015 | 0.01 ± 0.03 | 42.09 ± 0.13 | 48.4 |
|  | 0.820 ± 0.07 | 0.414 ± 0.018 | 0.2115±0.036 | 48.83 ± 0.15 | 51.6 |
| 200 K | 0.702±0.054 | 0.227±0.015 | 0.324±0.03 | 41.56±0.12 | 47.2 |
|  | 0.710 ± 0.06 | 0.414±0.0137 | 0.21456±0.027 | 48.14 ± 0.11 | 52.8 |
| 250 K | 0.66±0.062 | 0.169±0.018 | 0.628±0.036 | 39.84±0.14 | 49.2 |
|  | 0.665±0.063 | 0.341 ± 0.02 | -0.444±0.035 | 46.01 ± 0.14 | 50.8 |
| 300 K | 0.77 ± 0.03 | 0.17 ± 0.01 | 0.63 ± 0.02 | 37.4 ± 0.1 | 50.9 |
|  | 0.70 ± 0.03 | 0.32 ± 0.01 | -0.52 ± 0.02 | 44.4 ± 0.1 | 49.1 |

In a system with spiral magnetic order, the magneto-dielectricity arises due to changes in the spiral magnetic order due to the application of external magnetic field. The external magnetic field normally destroys the spiral order, thus decreasing the induced polarization ($P$) and increasing the polarization susceptibility ($\varepsilon'$), as per observed. In TbMnO$_3$, Kimura et al [11] argued that the Mn spin reorientation changes the exchange interaction energy and then brings about the lattice modulation owing to a finite spontaneous polarization. Following this argument, we suggest that in the SRT region the magnetic order can be considered to be 'frustrated', as over this temperature window, spins in two different directions are present as observed in our recent neutron diffraction studies on Ca$_2$Fe$_{1.2}$Al$_{0.8}$O$_5$ [30]. In the SRT region, the landscape of the sample possesses a distribution of spatial regions corresponding to magnetic phases with different spin directions. The boundaries between the different magnetic regions are expected to show a systematic variation in spin direction leading to the development of finite spontaneous polarization (via the Dzyaloshinskii-Moriya type spin orbit interaction). In a system which possesses strong spin lattice coupling, the lattice modulation is allied with the SRT and thus, the induced polarization is also expected to locally modulate [31,32]. In the SRT region, the external magnetic field also alters the spin configuration, in turn changing the induced polarization. Both these effects are reflected in the changes in $\varepsilon'$ of the sample around SRT and under the influence of magnetic field.

**Conclusions**

In conclusions, present studies give direct evidence of correlation among spin reorientation transition and positive magneto-dielectricity through strong spin-lattice coupling. The dielectric anomaly is observed across the spin reorientation transition signifying its link with the magnetism. In Ca$_2$FeCoO$_5$ compound it is shown by Mossbauer Spectroscopy that the relative area under the two sextets corresponding to octahedral and tetrahedral sites changes with temperature which has been interpreted as two distinct Debye temperatures corresponding to the octahedral and tetrahedral sites. The analysis of CS gave direct evidence of lattice instability across SRT. The strong spin-lattice coupling is reflected in modification of lattice parameters deduced from temperature dependent SXPD and anomalous behaviour of Raman modes across SRT. The magnetic field induced changes in spin dynamics across the SRT brings about the lattice modulation which in turn gives rise to the giant magneto-dielectricity. This study thus gives a new route to tune the magneto-dielectric response.

**Acknowledgement:** R. Rajamani is acknowledged for numerical analysis and Binoy K. De is acknowledge for help during SXRD measurements.